\documentclass[12pt]{revtex4}

\usepackage[cp1251]{inputenc}
\usepackage{graphics}
\usepackage{graphicx,epsfig}
\usepackage{epsfig}
\date{}

\newcommand\ba{\begin{eqnarray}}
\newcommand\ea{\end{eqnarray}}
\newcommand\nn{\nonumber}

\newcommand{\br}[1]{\left( #1 \right)}
\newcommand{\brs}[1]{\left[ #1 \right]}
\newcommand{\brf}[1]{\left\{ #1 \right\}}
\newcommand{\brm}[1]{\left| #1 \right|}
\newcommand{\TeV}{~\mbox{TeV}}
\newcommand{\GeV}{~\mbox{GeV}}

\newcommand{\mub} {~\mu\mbox{b}}
\newcommand{\nb} {~\mbox{nb}}
\newcommand{\pb} {~\mbox{pb}}

\makeatletter \@addtoreset{equation}{section} \makeatother

\renewcommand\thesubsection   {\thesection.\@arabic\c@subsection}

\begin{document}

\large

\title{Scalar meson production in proton-proton and proton-antiproton collisions}

\author{A.~I.~Ahmadov}
\email{ahmadov@theor.jinr.ru}
\affiliation{Joint Institute for Nuclear Research, Dubna, Russia}
\affiliation{Institute of Physics, Azerbaijan National Academy of Sciences, Baku, Azerbaijan}

\author{Yu.~M.~Bystritskiy}
\email{bystr@theor.jinr.ru}
\affiliation{Joint Institute for Nuclear Research, Dubna, Russia}

\author{E.~A.~Kuraev}
\email{kuraev@theor.jinr.ru}
\affiliation{Joint Institute for Nuclear Research, Dubna, Russia}

\begin{abstract}
Taking into account the exchange forces between protons of
scalar, pseudoscalar, vector and axial vector type the cross
sections of neutral and charged scalar mesons $a_0(980)$,
$a_+(980)$, $f_0(980)$, $\sigma(600)$ production are
calculated. The estimation for the facilities of moderately high
energies such as PANDA and NICA are presented. Similar analysis is
given for processes of charged and neutral Higgs boson production
at high energy proton-proton colliders such as Tevatron, RHIC
and LHC. The possible signal of neutral Higgs boson decay into two
oppositely charged leptons of different kinds is discussed.
Numerical estimations for parameters of modern colliders are given.
\end{abstract}

\maketitle

\section{Introduction}
\label{Introduction}

Proton-proton as well as proton-antiproton colliders of moderately
high energies with sufficiently high luminosities provide the
source of light scalar mesons. We investigate here the simplest mechanism
of meson creation which is the emission from the nucleon while the
exchange forces between the nucleons assumed to be
mediated by neutral scalar, pseudoscalar, vector and
axial-vector mesons. We introduce the nucleon-meson interaction
vertices as a product of coupling constants estimated in the
Nambu-Jona-Lasinio model \cite{NJL} and the relevant form factors.

For the case of large transfer momentum we chose the form of form
factors predicted by quark counting behavior \cite{Lepage}.

In such a facilities as NICA ($pp$ collisions with $s=20\GeV^2$)
and PANDA ($p \bar p$ collisions, $s \sim 30\GeV^2$) the cross-sections
of creation of scalar mesons
($a_0(980)$, $a_+(980)$, $f_0(980)$, $\sigma(600)$) for
large c.m. angles kinematics of $pp$-scattering is about $10^3\nb$.
These estimations was done in Sections~\ref{Amplitudes} and \ref{AppendixChiralAmplitudes}.

Neutral and charged Higgs bosons production in $pp$ collisions is
investigated in the same approach in Section~\ref{HiggsProduction}.
We assume the absence of formfactor suppression considering the
scattered proton to produce jet. In that case we estimate
the average cross section of the Higgs-boson (charged and neutral)
production on the level of several $\pb$ at the LHC energy range.

The possible signature of neutral Higgs boson production through
it's decay into two different lepton pairs was investigated in Section~\ref{HiggsSignature}.

The results are illustrated by the relevant spectral distributions.

\section{Differential cross sections of scalar meson production. Collinear approximation}
\label{Amplitudes}

Keeping in mind the smallness of meson masses compared with the
center of mass (c.m.) total energy of the initial nucleons
($M_S \sim M_P \sim M_V \sim M_A \sim M_p \ll \sqrt{s}$)
we can apply the chiral amplitudes method. Besides we will consider the
cross sections of production of scalar mesons with their emission in
the collinear kinematics.

Let us consider the following processes of neutral scalar meson production
\ba
    p\br{p_1, \lambda_1} + p\br{p_2, \lambda_2}
    \to
    p\br{p_1', \lambda_1'} + p\br{p_2', \lambda_2'}
    +
    S(k),\label{Process}
    \\
    p\br{p_1, \lambda_1} + \bar p\br{p_2, \lambda_2}
    \to
    p\br{p_1', \lambda_1'} + \bar p\br{p_2', \lambda_2'}
    +
    S(k),
\ea
where $S=\sigma(600)$, $f_0(980)$, $a_0(980)$ is the neutral scalar
meson.
In case of charged scalar meson $a_+(980)$ production one
of proton in the final state must be replaced by the neutron.

The sufficiently high energies of colliders allows to neglect
masses of all particles and use the chiral amplitudes method \cite{Berends}.

The collinear kinematics of scalar meson emission dominates.
Using the "quasi-real electron" method \cite{BFK}
we obtain for process (\ref{Process}) for scalar meson emission from initial
nucleons:
\ba
    M^{\lambda_1 \lambda_1' \lambda_2 \lambda_2'}
    &=&
    \sum_{r=\pm 1}
    M_B^{r \lambda_1' \lambda_2 \lambda_2'}
    \br{p_1-k,p_2;p_1',p_2'}
    \frac{\bar u^r\br{p_1-k} u^{\lambda_1}\br{p_1}}{-2\br{p_1 k}}
    g_S + \nn\\
    &+&
    \sum_{r=\pm 1}
    M_B^{\lambda_1 \lambda_1' r \lambda_2'}
    \br{p_1,p_2-k;p_1',p_2'}
    \frac{\bar u^r\br{p_2-k} u^{\lambda_2}\br{p_2}}{-2\br{p_2 k}}
    g_S,
\ea
for the case of emission from final nucleons we obtain
\ba
    M^{\lambda_1 \lambda_1' \lambda_2 \lambda_2'}
    &=&
    \sum_{r=\pm 1}
    M_B^{\lambda_1 r \lambda_2 \lambda_2'}
    \br{p_1,p_2;p_1'+k,p_2'}
    \frac{\bar u^{\lambda_1'}\br{p_1'} u^r\br{p_1'+k}}{2\br{p_1' k}}
    g_S + \nn\\
    &+&
    \sum_{r=\pm 1}
    M_B^{\lambda_1 \lambda_1' \lambda_2 r}
    \br{p_1,p_2;p_1',p_2'+k}
    \frac{\bar u^{\lambda_2'}\br{p_2'} u^r\br{p_2'+k}}{2\br{p_2' k}}
    g_S,
\ea
where $M_B^{\lambda_1 \lambda_1' \lambda_2 \lambda_2'}$
is the matrix element of $pp \to pp$ subprocess in Born approximation.

Chiral states of nucleon and antinucleon are defined as \cite{Berends}
\ba
u^\lambda(p)=\omega_\lambda u(p), \qquad v^\lambda(p)=\omega_{-\lambda} v(p),
\ea
\ba
\omega_\lambda=\frac{1}{2}\br{1+\lambda\gamma_5}, \qquad \lambda=\pm. \nn
\ea

The relevant cross sections then have a form (center of mass of
initial particles implied):
\ba
    \frac{d\sigma^{\lambda_1 \lambda_1' \lambda_2 \lambda_2'}}{dC dx}
    &=&
    \frac{g_S^2\br{1-x}L(x)}{64 \pi^3 s \br{2-x\br{1-C}}^2}
    \brm{M_B^{-\lambda_1 \lambda_1' \lambda_2 \lambda_2'}}^2
    ,\qquad \mbox{when~} \vec k || \vec p_1,
    \label{ScalarCS1}
    \\
    \frac{d\sigma^{\lambda_1 \lambda_1' \lambda_2 \lambda_2'}}{dC dx}
    &=&
    \frac{g_S^2\br{1-x}L(x)}{64 \pi^3 s \br{2-x\br{1+C}}^2}
    \brm{M_B^{\lambda_1 \lambda_1' -\lambda_2 \lambda_2'}}^2
    ,\qquad  \mbox{when~} \vec k || \vec p_2,
    \label{ScalarCS2}
    \\
    \frac{d\sigma^{\lambda_1 \lambda_1' \lambda_2 \lambda_2'}}{dC dx}
    &=&
    \frac{g_S^2\br{1-x} L(x)}{256 \pi^3 s}
    \brm{M_B^{\lambda_1 -\lambda_1' \lambda_2 \lambda_2'}}^2
    , \qquad \mbox{when~} \vec k || \vec p_1',
    \label{ScalarCS3}
    \\
    \frac{d\sigma^{\lambda_1 \lambda_1' \lambda_2 \lambda_2'}}{dC dx}
    &=&
    \frac{g_S^2\br{1-x} L(x)}{256 \pi^3 s}
    \brm{M_B^{\lambda_1 \lambda_1' \lambda_2 -\lambda_2'}}^2
    , \qquad \mbox{when~} \vec k || \vec p_2',
    \label{ScalarCS4}
\ea
where $x = k_0/E$, where $k_0$ and $E$ are
the energies of created scalar meson and initial proton
correspondingly. $C=\cos\br{\theta}$ and $\theta$ is the angle between direction of
initial proton beam direction $\vec p_1$ and scattered proton
$\vec p'_1$. The factor $L(x)$ is the large logarithm which
enhances the probability of meson production in collinear
kinematics:
\ba
    L(x) =
    \ln\br{
        \frac{4E^2 x^2}{M_p^2 x^2 + M_S^2}
    },
\ea
where $M_p$ and $M_S$ are the masses of proton and the created scalar meson
correspondingly.

The kinematical invariants we define in the following manner
\ba
    s=2\br{p_1p_2}=4E^2, \quad
    t = -2\br{p_1p_1'} =-\frac{s}{2}\br{1-C}, \nn
\ea
\ba
    u = -\frac{s}{2}\br{1+C}, \qquad s+t+u = 0. \nn
\ea

In case of colliders of moderate energies the collinear approximation allows to evaluate the
cross section with the accuracy of order 10\%.

In the next section we will evaluate the Born amplitudes
$M_B^{\lambda_1 \lambda_1' \lambda_2 \lambda_2'}$ from
(\ref{ScalarCS1}), (\ref{ScalarCS2}), (\ref{ScalarCS3}), (\ref{ScalarCS4}).

\section{Chiral amplitudes of $pp \to pp$ subprocess in Born approximation}
\label{AppendixChiralAmplitudes}

Matrix element of the subprocess
\ba
    p\br{p_1, \lambda_1} + p\br{p_2, \lambda_2}
    \to
    p\br{p_1', \lambda_1'} + p\br{p_2', \lambda_2'}
\ea
have a form:
\ba
    M_B^{\lambda_1 \lambda_1' \lambda_2 \lambda_2'}
    &=&
    \sum_{i=S,P,V,A}
    \frac{F_i^2\br{t}}{t}
    \brs{\bar u^{\lambda_1'}\br{p_1'} \Gamma_i \omega_{\lambda_1} u\br{p_1}}
    \brs{\bar u^{\lambda_2'}\br{p_2'} \Gamma_i \omega_{\lambda_2} u\br{p_2}}
    - \nn\\
    &-&
    \sum_{i=S,P,V,A}
    \frac{F_i^2\br{u}}{u}
    \brs{\bar u^{\lambda_2'}\br{p_2'} \Gamma_i \omega_{\lambda_1} u\br{p_1}}
    \brs{\bar u^{\lambda_1'}\br{p_1'} \Gamma_i \omega_{\lambda_2} u\br{p_2}},
    \label{BornSubprocess}
\ea
where $\Gamma_i$ are the vertexes of interaction of the scalar, pseudoscalar, vector
and axial-vector mesons with the protons:
\ba
    \Gamma_S = 1, \qquad
    \Gamma_P = \gamma_5, \qquad
    \Gamma_V = \gamma_\mu, \qquad
    \Gamma_A = \gamma_\mu \gamma_5.
\ea

Meson-proton form factors are model-dependent ones. At zero value
of momentum transferred they are determined by the relevant
current-algebra values. The behavior at large momentum is
determined by quark-counting rule:
\ba
F_{i}(t)&=&\frac{g_i}{\br{1+\frac{-t}{M_0^2}}^2},
\qquad
i = S, P, V, A, \nn\\
F_{i}(u)&=&\frac{g_i}{\br{1+\frac{-u}{M_0^2}}^2}, \qquad
Re\brs{F_{i}(s)}=\frac{g_i}{\br{1+\frac{s}{M_0^2}}^2}, \nn
\ea
where $M_0 \sim 1-2 \GeV$ is the mass scale characterizing the region the
quark-counting approximation is valid. The vertex constants $g_i$
we take following the prescription of Nambu-Jona-Lazinio model
\cite{NJL}:
\ba
    g_S \approx 10, \qquad
    g_P \approx 12, \qquad
    g_V = g_A \approx 6.
\ea
Inserting the projection operators
\ba
    {\cal P}_1 = \frac{N_1}{N_1} = 1,
    \qquad
    N_1 = \brs{\bar u\br{p_1} \hat p_1' \omega_+ u\br{p_1}}
          \brs{\bar u\br{p_2} \hat p_2' \omega_+ u\br{p_1'}},
    \\
    {\cal P}_2 = \frac{N_2}{N_2} = 1,
    \qquad
    N_2 = \brs{\bar u\br{p_1} \hat p_2' \omega_+ u\br{p_1'}}
          \brs{\bar u\br{p_2} \hat p_1' \omega_+ u\br{p_2'}},
\ea
into (\ref{BornSubprocess}) and using the completeness condition
of type $u^\lambda\br{p} \bar u^\lambda\br{p} = \omega_\lambda \hat p$,
we obtain for $M_B^{++++}$:
\ba
    M_B^{++++} &=& \frac{S_1}{N_1}\frac{1}{t} - \frac{S_2}{N_2}\frac{1}{u},
\ea
\ba
    S_1 = 2 t s^2 \br{F_V^2\br{t} + F_A^2\br{t}},
    \quad
    S_2 = 2 u s^2 \br{F_V^2\br{u} + F_A^2\br{u}}. \nn
\ea
Using further the explicit values of $N_{1,2}$:
\ba
    \brm{N_1}^2 = \br{s t}^2,
    \qquad
    \brm{N_2}^2 = \br{s u}^2,
    \qquad
    N_1 N_2^* = -s^2 t u,
\ea
we obtain for $\brm{M_{pp}^{++++}}^2$:
\ba
    \brm{M_{pp}^{++++}}^2 &=& 4s^2
    \brm{\frac{F_V^2\br{t} + F_A^2\br{t}}{t} + \frac{F_V^2\br{u} + F_A^2\br{u}}{u}}^2.
\ea
Performing the similar algebraic manipulations we calculate the
remaining non-vanishing chiral 
amplitudes $\brm{M_{pp}^{\lambda_1 \lambda_1' \lambda_2 \lambda_2'}}^2$:
\ba
    \brm{M_{pp}^{++--}}^2 &=&
    \brm{\frac{2u}{t}\br{F_V^2\br{t} - F_A^2\br{t}} - \br{F_S^2\br{u} - F_P^2\br{u}}}^2,
    \nn\\
    \brm{M_{pp}^{+--+}}^2 &=&
    \brm{\br{F_S^2\br{t} - F_P^2\br{t}} - \frac{2t}{u}\br{F_V^2\br{u} - F_A^2\br{u}}}^2,
    \\
    \brm{M_{pp}^{+-+-}}^2 &=&
    \brm{F_S^2\br{t} + F_P^2\br{t} + F_S^2\br{u} + F_P^2\br{u}}^2. \nn
\ea
Similar calculation can be done for case of proton-antiproton collision
$\brm{M_{p \bar p}^{\lambda_1 \lambda_1' \lambda_2 \lambda_2'}}^2$:
\ba
    \brm{M_{p \bar p}^{++++}}^2 &=&
    \brm{\frac{2s}{t} \br{F_V^2\br{t} - F_A^2\br{t}} - \br{F_S^2\br{s} - F_P^2\br{s}}}^2,
    \nn\\
    \brm{M_{p \bar p}^{++--}}^2 &=& 4u^2
    \brm{\frac{F_V^2\br{t} + F_A^2\br{t}}{t} + \frac{F_V^2\br{s} + F_A^2\br{s}}{s}}^2,
    \\
    \brm{M_{p \bar p}^{+--+}}^2 &=&
    \brm{F_S^2\br{t} - F_P^2\br{t} + F_S^2\br{s} - F_P^2\br{s}}^2,
    \nn\\
    \brm{M_{p \bar p}^{+-+-}}^2 &=&
    \brm{\br{F_S^2\br{t} + F_P^2\br{t}} - \frac{2t}{s}\br{F_V^2\br{s} + F_A^2\br{s}}}^2. \nn
\ea
In case of proton-neutron and antiproton-neutron scattering we obtain:
\ba
    \brm{M_{p n}^{++++}}^2 &=& \frac{4s^2}{t^2}
    \brm{F_V^2\br{t} + F_A^2\br{t}}^2,
    \\
    \brm{M_{p n}^{++--}}^2 &=& \frac{4u^2}{t^2}
    \brm{F_V^2\br{t} - F_A^2\br{t}}^2,
    \nn\\
    \brm{M_{p n}^{+--+}}^2 &=&
    \brm{F_S^2\br{t} - F_P^2\br{t}}^2,
    \nn\\
    \brm{M_{p n}^{+-+-}}^2 &=&
    \brm{F_S^2\br{t} + F_P^2\br{t}}^2,
    \nn\\
    \brm{M_{\bar p n}^{++++}}^2 &=& \frac{4s^2}{t^2}
    \brm{F_V^2\br{t} - F_A^2\br{t}}^2,
    \\
    \brm{M_{\bar p n}^{++--}}^2 &=& \frac{4u^2}{t^2}
    \brm{F_V^2\br{t} + F_A^2\br{t}}^2,
    \nn\\
    \brm{M_{\bar p n}^{+--+}}^2 &=&
    \brm{F_S^2\br{t} - F_P^2\br{t}}^2,
    \nn\\
    \brm{M_{\bar p n}^{+-+-}}^2 &=&
    \brm{F_S^2\br{t} + F_P^2\br{t}}^2.
    \nn
\ea

Due to parity conservation we have $|M^{\lambda}|^2=|M^{-\lambda}|^2$,
so we can choose $\lambda_1=+$.

\begin{figure}
    \includegraphics[width=0.8\textwidth]{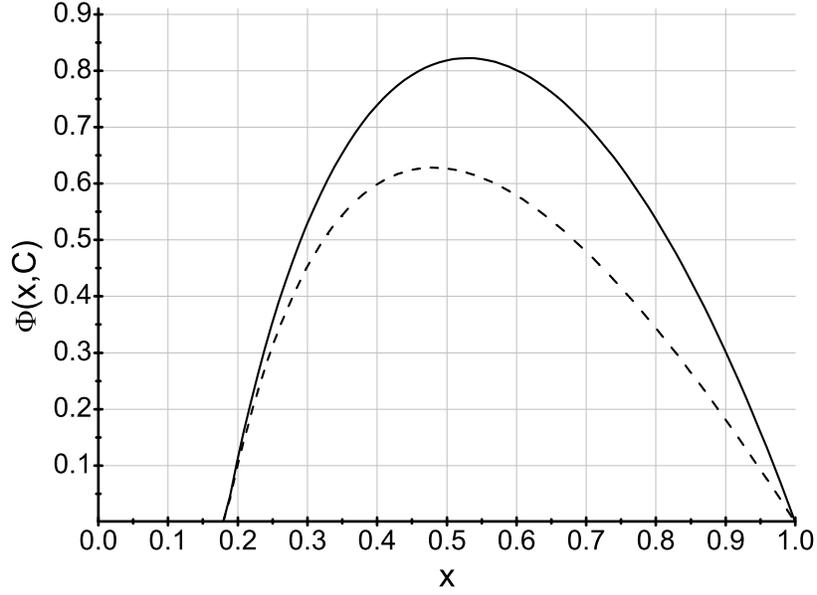}
    \caption{The value $\Phi\br{x}$ for production of $f_0(980)$ meson
    in case when angle between the momenta of initial and scattered protons
    is $\theta=60^o$ (where $C=\cos\br{\theta}$.
    $x=2 E_{f_0}/\sqrt{s}$ is the energy fraction of $f_0$ meson.
    The solid line is the case then $f_0$ meson is produced along
    $\vec p_1$ (i.e. $d\sigma^{-+++}(\vec k || \vec p_1)$),
    dashed line is the case then $f_0$ meson is produced along
    $\vec p_1'$
    (i.e. $d\sigma^{-+++}(\vec k || \vec p_1')$).
    \label{Fig60pppp}}
\end{figure}

\begin{figure}
    \includegraphics[width=0.8\textwidth]{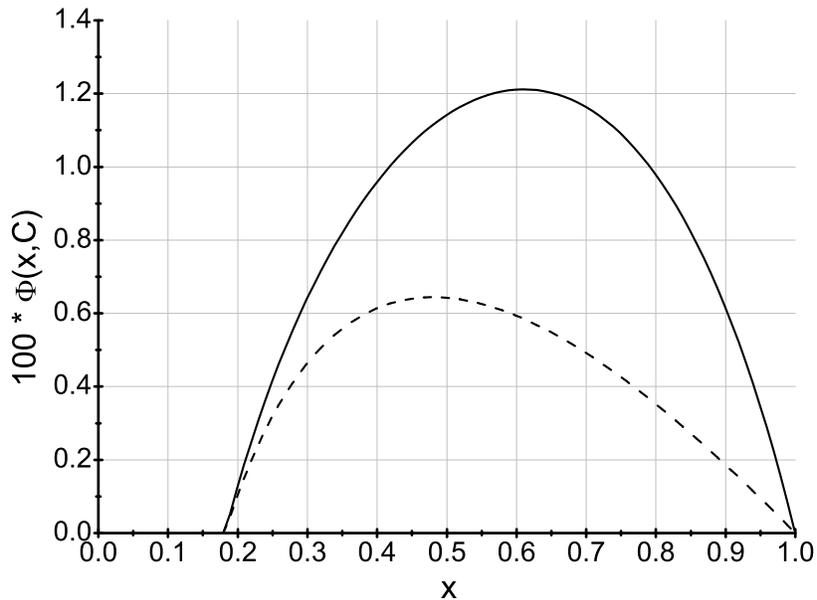}
    \caption{The notations are the same as in Fig.~\ref{Fig60pppp}.
    The case when $\theta=90^o$.
    \label{Fig90pppp}}
\end{figure}

\begin{figure}
    \includegraphics[width=0.8\textwidth]{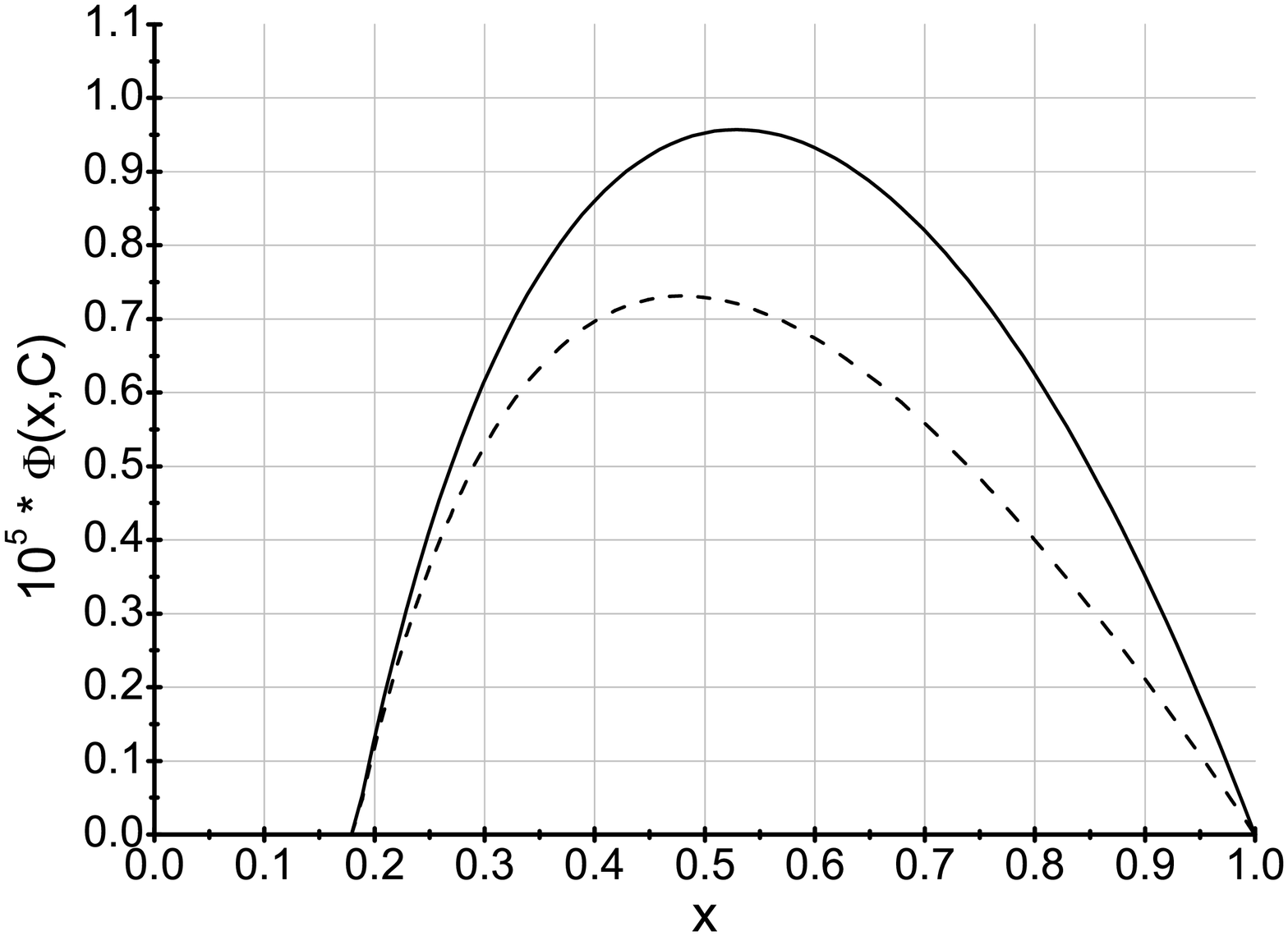}
    \caption{The notations are the same as in Fig.~\ref{Fig60pppp}.
    The case when $\theta=60^o$ and the spiral amplitude spin state
    is $\brf{\lambda} = \brf{-+--}$.
    \label{Fig60ppmm}}
\end{figure}

\begin{figure}
    \includegraphics[width=0.8\textwidth]{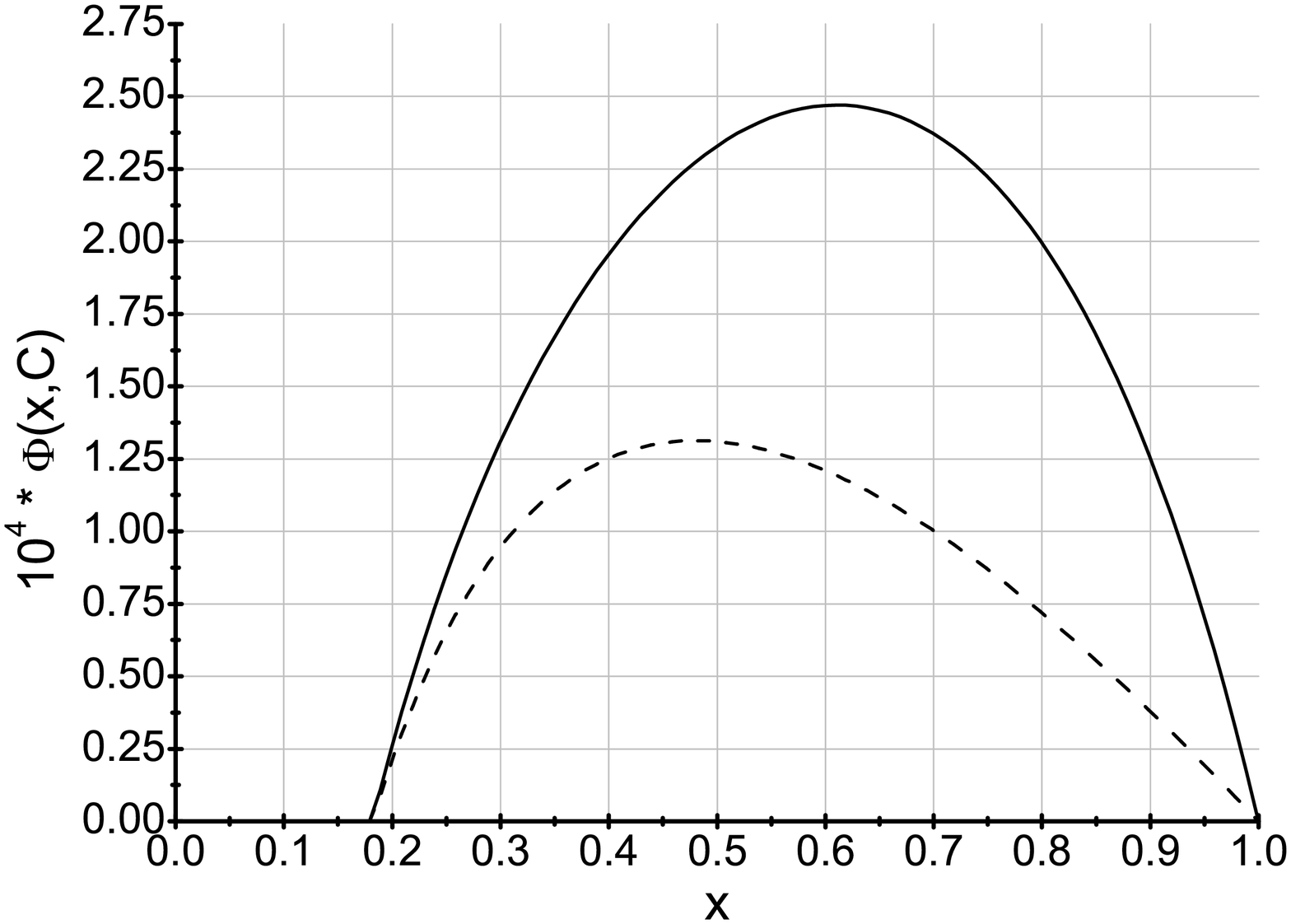}
    \caption{The notations are the same as in Fig.~\ref{Fig60pppp}.
    The case when $\theta=90^o$ and the spiral amplitude spin state
    is $\brf{\lambda} = \brf{-+--}$.
    \label{Fig90ppmm}}
\end{figure}

The spectral distributions
\ba
    \frac{1}{\sigma_0}\frac{d\sigma}{dc dx}=\Phi(x,c),
    \qquad
    \sigma_0 = \frac{g_s^2}{64 \pi^3 s}=\frac{20 \mub}{s(\GeV^2)},
\ea
for several fixed values of $C=\cos\br{\theta}$ are presented in
Fig.~\ref{Fig60pppp}, \ref{Fig90pppp}, \ref{Fig60ppmm}, \ref{Fig90ppmm}.

\section{Neutral and charged Higgs boson production in $p p$ collisions}
\label{HiggsProduction}

Consider the simplest channel of neutral Higgs boson production at
proton-proton collision:
\ba
p(p_1)+p(p_2)\to p(p_1')+p(p_2')+H(k),
\ea
\ba
s=(p_1+p_2)^2 \gg k^2=M^2 \gg M_p^2. \nn
\ea
We suppose Higgs boson interact coherently with quarks of a
proton, resulting in the vertex $g_H$ of Higgs boson-proton-proton
interaction of a form \cite{Amsler:2008zzb}:
\ba
g_H=\frac{M_p}{v}, \qquad v=256 \GeV.
\ea

\begin{figure}
    \includegraphics[width=0.8\textwidth]{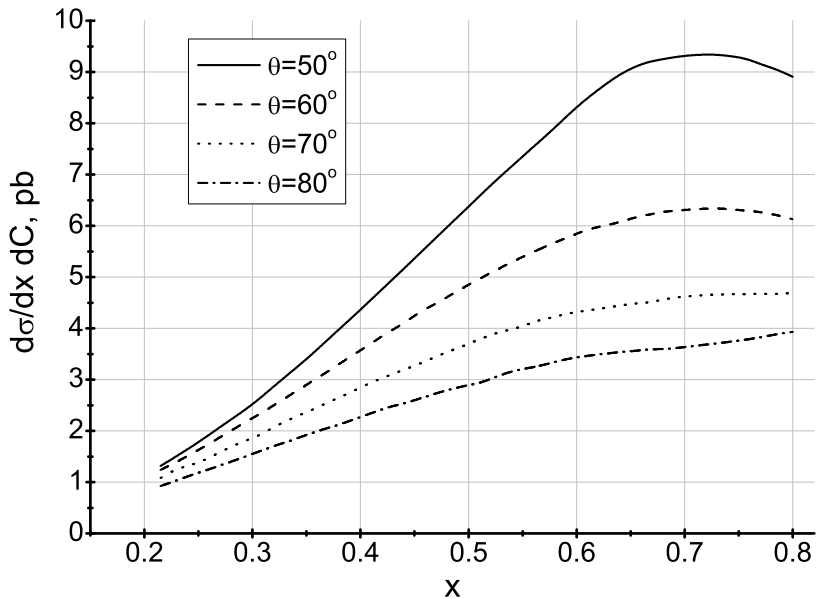}
    \caption{\label{FigHiggs}
    The cross section for production of Higgs boson in $pp$ collisions
    at $\sqrt{s} = 14\TeV$
    as a function of Higgs energy fraction $x=2 \omega/\sqrt{s}$.
    Different curves correspond to different angle of between the initial beam direction
    and the direction of produces Higgs boson momenta $C = \cos\br{\theta}$.
    Non-complanarity condition is $D > 0.01$.}
\end{figure}

Taking into account the identity of protons in the final state only $8$ tree type Feynman
diagrams must be considered. The relevant contributions schematically can be written as
\ba
g_H{\cal M} &=& g_H \brs{M_1+M_2+M_3+M_4}, \nn
\ea
\ba
M_1 &=& \sum_{i=S,P,V,A} F_i^2(t_1)\frac{\bar{u}(p_2')\Gamma_i u(p_2)}{t_1}\bar u(p_1')O_1^i u(p_1),
\quad
O_1^i=-\frac{\Gamma_i \hat{k}}{\chi_1}+\frac{\hat{k}\Gamma_i}{\chi_1'} ; \nn \\
M_2 &=& \sum_{i=S,P,V,A} F_i^2(t) \frac{\bar{u}(p_1')\Gamma_i u(p_1)}{t}\bar u(p_2')O_2^i u(p_2),
\quad
O_2^i=-\frac{\Gamma_i \hat{k}}{\chi_2}+\frac{\hat{k}\Gamma_i}{\chi_2'} ; \nn \\
M_3 &=& -\sum_{i=S,P,V,A} F_i^2(u_1) \frac{\bar{u}(p_1')\Gamma_i u(p_2)}{u_1}\bar u(p_2')O_3^i u(p_1),
\quad
O_3^i=-\frac{\Gamma_i \hat{k}}{\chi_1}+\frac{\hat{k}\Gamma_i}{\chi_2'} ; \nn \\
M_4 &=& -\sum_{i=S,P,V,A} F_i^2(u) \frac{\bar{u}(p_2')\Gamma_i
u(p_1)}{u}\bar u(p_1')O_4^i u(p_2), \,\, O_4^i=-\frac{\Gamma_i
\hat{k}}{\chi_2}+\frac{\hat{k}\Gamma_i}{\chi_1'}, \nn
\ea
with the kinematical invariants now defined as
\ba
s=2p_1p_2, \qquad  t=-2p_1p_1', \qquad u=-2p_1p_2', \nn
\ea
\ba
s_1=2p_1'p_2', \qquad t_1=-2p_2p_2', \qquad u_1=-2p_2p_1', \nn
\ea
\ba
\chi_i=2p_i k, \qquad \chi'=2p_i' k, \qquad p_i^2=p_i^{'2}=0, \quad i=1,2; \quad k^2=M_H^2,  \nn
\ea
\ba
s+s_1+t+t_1+u+u_1=M_H^2,
\ea
indexes $S$, $P$, $V$, $A$ denote the type of exchange forces between the nucleons.
With the choice $\lambda_1=+1$ only 4 chiral amplitudes entering in the
matrix element squared given above are nonzero:
\ba
{\cal M}^{+++-}, \quad {\cal M}^{++-+}, \quad {\cal M}^{+-++}, \quad {\cal M}^{+---}.\nn
\ea
In terms of $M_i$ they are:
\ba
{\cal M}^{+++-}=M_1^{s,p}+M_2^{v,a}+M_3^{v,a}+M_4^{s,p}; \nn \\
{\cal M}^{+---}=M_1^{v,a}+M_2^{s,p}+M_3^{v,a}+M_4^{s,p}; \nn \\
{\cal M}^{+-++}=M_1^{v,a}+M_2^{s,p}+M_3^{s,p}+M_4^{v,a}; \nn \\
{\cal M}^{++-+}=M_1^{s,p}+M_2^{v,a}+M_3^{s,p}+M_4^{v,a}. \nn
\ea
Now we use again the projection operators
\ba
P_{12}=\frac{N_{12}}{N_{12}}, \qquad P_{34}=\frac{N_{34}}{N_{34}}.
\ea
We can choose the quantities $N_{12}$, $N_{34}$ for each chiral state
to provide their equality
\ba
|N_{12}|^2_{\{\lambda\}}=|N_{34}|^2_{\{\lambda\}}=N_{12}^{\{\lambda\}}N_{34}^{*\{\lambda\}}
\equiv N^{\{\lambda\}},
\ea
where
$\{\lambda\}=\{\lambda_1\lambda_1'\lambda_2\lambda_2'\}$. Using
the completeness condition for the chiral states of protons
$u_\lambda(p)\bar {u}_\lambda (p)= \omega_\lambda\hat{p}$ chiral
amplitudes can be written in form 
\ba {\cal M}^{\{\lambda\}}=
\brs{\frac{1}{N_{12}}\br{\frac{S_1}{t_1}+\frac{S_2}{t}}-
\frac{1}{N_{34}}\br{\frac{S_3}{u_1}+\frac{S_4}{u}}}_{\{\lambda\}}.
\ea
And the matrix element square then reads as
\ba
|{\cal
M}^{\{\lambda\}}|^2=\frac{1}{N^{\{\lambda\}}}\brm{\frac{S_1}{t_1}+\frac{S_2}{t}-
\frac{S_3}{u_1}-\frac{S_4}{u}}_{\{\lambda\}}^2.
\label{MSquare}
\ea
Below we present the explicit form of the values from (\ref{MSquare}) for
definite choice of $\{\lambda\}=\{\lambda_1\lambda_1'\lambda_2\lambda_2'\}$.
In case then $\{\lambda_1\lambda_1'\lambda_2\lambda_2'\}=\{+++-\}$
we have:
\ba
N_{12}&=&\brs{\bar{u}_2\hat{p}'_2\omega_+ u'_1}
\brs{\bar{u}_1\omega_- u'_2}; \qquad
N^{+++-}=s_1 t_1 u; \nn\\
N_{34}&=&\brs{\bar{u}_2\omega_- u'_1}\brs{\bar{u}_1\hat{p}'_2\omega_+ u'_1}; \nn \\
S_1 &=& F_S^2(t_1)~Tr\brs{\hat{p}'_2\Gamma_S\hat{p}_2\hat{p}'_2\hat{p}'_1O_1^{S}\hat{p}_1\omega_-}
     +  F_P^2(t_1)~Tr\brs{\hat{p}'_2 \Gamma_P\hat{p}_2\hat{p}'_2\hat{p}'_1O_1^{P}\hat{p}_1\omega_-}; \nn \\
S_2 &=& F_V^2(t)~Tr\brs{\hat{p}'_1\Gamma_V\hat{p}_1\hat{p}'_2O_2^{V}\hat{p}_2\hat{p}'_2\omega_+}
     +  F_A^2(t)~Tr\brs{\hat{p}'_1\Gamma_A\hat{p}_1\hat{p}'_2O_2^{A}\hat{p}_2\hat{p}'_2\omega_+}; \nn \\
S_3 &=& F_V^2(u_1)~Tr\brs{\hat{p}'_1\Gamma_{V}\hat{p}_2\hat{p}'_2O_3^{V}\hat{p}_1\hat{p}'_2\omega_+}
     +  F_A^2(u_1)~Tr\brs{\hat{p}'_1\Gamma_{A}\hat{p}_2\hat{p}'_2O_3^{A}\hat{p}_1\hat{p}'_2\omega_+}; \nn \\
S_4 &=& F_S^2(u)~Tr\brs{\hat{p}'_2\Gamma_{S}\hat{p}_1\hat{p}'_2\hat{p}'_1O_4^{S}\hat{p}_2\omega_-}
     +  F_P^2(u)~Tr\brs{\hat{p}'_2\Gamma_{P}\hat{p}_1\hat{p}'_2\hat{p}'_1O_4^{P}\hat{p}_2\omega_-}. \nn
\ea
For chiral state $\{+---\}$ we have
\ba
N_{12} &=&
\brs{\bar{u}_2\hat{p}_1\omega_- u'_1}\brs{\bar{u}_1\omega_- u_2'};
\qquad
N^{+---}=s t u; \nn \\
N_{34} &=& \brs{\bar{u}_2\hat{p}_1\omega_- u'_2}\brs{\bar{u}_1\omega_- u'_1}; \nn \\
S_1 &=& F_V^2(t_1)~Tr\brs{\hat{p}'_2\Gamma_{V}\hat{p}_2\hat{p}_1\hat{p}'_1O_1^{V}\hat{p}_1\omega_-}
     +  F_A^2(t_1)~Tr\brs{\hat{p}'_2\Gamma_{A}\hat{p}_2\hat{p}_1\hat{p}'_1O_1^{A}\hat{p}_1\omega_-}; \nn \\
S_2 &=& F_S^2(t)~Tr\brs{\hat{p}'_1\Gamma_{S}\hat{p}_1\hat{p}'_2O_2^{S}\hat{p}_2\hat{p}_1\omega_-}
     +  F_P^2(t)~Tr\brs{\hat{p}'_1\Gamma_{P}\hat{p}_1\hat{p}'_2O_2^{P}\hat{p}_2\hat{p}_1\omega_-}; \nn \\
S_3 &=& F_V^2(u_1)~Tr\brs{\hat{p}'_1\Gamma_{V}\hat{p}_2\hat{p}_1\hat{p}'_2O_3^{V}\hat{p}_1\omega_-}
     +  F_A^2(u_1)~Tr\brs{\hat{p}'_1\Gamma_{A}\hat{p}_2\hat{p}_1\hat{p}'_2O_3^{A}\hat{p}_1\omega_-}; \nn \\
S_4 &=& F_S^2(u)~Tr\brs{\hat{p}'_2\Gamma_{S}\hat{p}_1\hat{p}'_1O_4^{S}\hat{p}_2\hat{p}_1\omega_-}
     +  F_P^2(u)~Tr\brs{\hat{p}'_2\Gamma_{P}\hat{p}_1\hat{p}'_1O_4^{P}\hat{p}_2\hat{p}_1\omega_-}. \nn
\ea
For chiral state $\{++-+\}$ we have
\ba
N_{12} &=&
\brs{\bar{u}_2\omega_+ u'_1}\brs{\bar{u}_1\hat{p}_2\omega_+ u'_2};
\qquad
N^{++-+}=s t_1 u_1; \nn \\
N_{34} &=& \brs{\bar{u}_2\omega_+ u'_2}\brs{\bar{u}_1\hat{p}_2\omega_+ u'_1}; \nn \\
S_1 &=& F_S^2(t_1)~Tr\brs{\hat{p}'_2\Gamma_{S}\hat{p}_2\hat{p}'_1O_1^{S}\hat{p}_1\hat{p}_2\omega_+}
     +  F_P^2(t_1)~Tr\brs{\hat{p}'_2\Gamma_{P}\hat{p}_2\hat{p}'_1O_1^{P}\hat{p}_1\hat{p}_2\omega_+}; \nn \\
S_2 &=& F_V^2(t)~Tr\brs{\hat{p}'_1\Gamma_{V}\hat{p}_1\hat{p}_2\hat{p}'_2O_2^{V}\hat{p}_2\omega_+}
     +  F_A^2(t)~Tr\brs{\hat{p}'_1\Gamma_{A}\hat{p}_1\hat{p}_2\hat{p}'_2O_2^{A}\hat{p}_2\omega_+}; \nn \\
S_3 &=& F_S^2(u_1)~Tr\brs{\hat{p}'_1\Gamma_{S}\hat{p}_2\hat{p}'_2O_3^{S}\hat{p}_1\hat{p}_2\omega_+}
     +  F_P^2(u_1)~Tr\brs{\hat{p}'_1\Gamma_{P}\hat{p}_2\hat{p}'_2O_3^{P}\hat{p}_1\hat{p}_2\omega_+}; \nn \\
S_4 &=& F_V^2(u)~Tr\brs{\hat{p}'_2\Gamma_{V}\hat{p}_1\hat{p}_2\hat{p}'_1O_4^{V}\hat{p}_2\omega_+}
     +  F_A^2(u)~Tr\brs{\hat{p}'_2\Gamma_{A}\hat{p}_1\hat{p}_2\hat{p}'_1O_4^{A}\hat{p}_2\omega_+}. \nn
\ea
For chiral state $\{+-++\}$ we have
\ba
N_{12} &=&
\brs{\bar{u}_2\omega_- u'_1}\brs{\bar{u}_1\hat{p}'_1\omega_+ u_2'};
\qquad
N^{+-++}=s_1 t u_1; \nn \\
N_{34} &=& \brs{\bar{u}_2\hat{p}'_1\omega_+ u'_2}\brs{\bar{u}_1\omega_- u'_1}; \nn \\
S_1 &=& F_V^2(t_1)~Tr\brs{\hat{p}'_2\Gamma_V\hat{p}_2\hat{p}'_1O_1^V\hat{p}_1\hat{p}'_1\omega_+}
     +  F_A^2(t_1)~Tr\brs{\hat{p}'_2\Gamma_A\hat{p}_2\hat{p}'_1O_1^A\hat{p}_1\hat{p}'_1\omega_+}; \nn \\
S_2 &=& F_S^2(t)~Tr\brs{\hat{p}'_1\Gamma_S\hat{p}_1\hat{p}'_1\hat{p}'_2O_2^S\hat{p}_2\omega_-}
     +  F_P^2(t)~Tr\brs{\hat{p}'_1g^\Gamma_P\hat{p}_1\hat{p}'_1\hat{p}'_2O_2^P\hat{p}_2\omega_-}; \nn \\
S_3 &=& F_S^2(u_1)~Tr\brs{\hat{p}'_1\Gamma_S\hat{p}_2\hat{p}'_1\hat{p}'_2O_3^S\hat{p}_1\omega_-}
     +  F_P^2(u_1)~Tr\brs{\hat{p}'_1\Gamma_P\hat{p}_2\hat{p}'_1\hat{p}'_2O_3^P\hat{p}_1\omega_-}; \nn \\
S_4 &=& F_V^2(u)~Tr\brs{\hat{p}'_2\Gamma_V\hat{p}_1\hat{p}'_1\hat{p}'_2O_4^V\hat{p}_2\omega_-}
     +  F_A^2(u)~Tr\brs{\hat{p}'_2\Gamma_A\hat{p}_1\hat{p}'_1\hat{p}'_2O_4^A\hat{p}_2\omega_-}. \nn
\ea
For the case of the charged Higgs boson production we must put $S_3=S_4=0$
in (\ref{MSquare}).
The cross section of Higgs meson production have a form
\ba
    \frac{1}{\sigma_H}\frac{d\sigma_{H}^{\{\lambda\}}}{dC dx}
    &=&
    \frac{dx' dC_0}{\sqrt{D\br{C_0,C,C'}}}
    \brm{M^{\{\lambda\}}}^2 s, \\
    \sigma_H &=& \frac{g_H^2}{512 \pi^4 s} = \frac{0.13\pb}{s\br{\GeV^2}}, \nn
\ea
where $x=\omega/E$, $x'=E_1'/E$, $\sigma=M_H^2/s$,
$C_0 = \cos\br{\vec p_1,\vec p_1'}$,
$C = \cos\br{\vec k,\vec p_1}$, and
\ba
    C' &=& 1 + \frac{2}{x x'} \br{1-x-x'+\sigma}, \nn\\
    D\br{C_0,C,C'} &=&
    1 - C^2 - C'^2 - C_0^2 + 2 C C' C_0,
    \qquad D>0. \nn
\ea
The relevant parametrization of kinematical invariants using the conservation laws
can be written as:
\ba
\chi_1=\frac{s x}{2}(1-C), \chi_2=\frac{s x}{2}(1+C),\chi'_1=s(1-x_2-\sigma) \nn\\
t=-\frac{s x'}{2}(1-C_0); u_1=-\frac{s x'}{2}(1+C_0); \chi'_2=s(1-x'-\sigma); \nn\\
t_1=t+\chi'_1-\chi_1+M_H^2; u=u_1+\chi'_1-\chi_2, x=\frac{\omega}{E}, \nn\\
x'=\frac{E'_1}{E}, \,\, x_2=\frac{E'_2}{E},\,\, s=4E^2. \ea
We will argue below that estimating the Higgs boson production we
can replace $F_i(t) =g_i$ (see Discussion). 
To make some numerical illustration we present the
 spectral distributions over
produced Higgs boson energy fraction with fixed angles of produces Higgs boson momenta
\ba
\frac{1}{\sigma_H}
\frac{d\sigma}{dC dx}=\Phi_H\br{C,x} = \int\limits_{-1}^1 dC_0
\int\limits_{x_0} ^1 dx' \cdot \frac{s \cdot
|M^{\lambda}|^2}{\sqrt D}, \qquad D>0.
\ea
Keeping in mind the condition of possibility to measure the final particles
(we avoid collinear and planar kinematics) we put
additional constrains: $D > 0.01$.
The mass of Higgs boson we took as $M_{H_0} = M_{H_+} = 140\GeV$.
The spectral distributions $\Phi_H\br{C,x}$ for neutral Higgs boson production
within these assumptions
for several values of Higgs boson emission angle $C=\cos(\theta)$,
$\theta=50^o-80^o$ are presented in Fig.~\ref{FigHiggs}.
The obtained results for neutral Higgs boson production are in
the satisfactory agreement with estimations given in \cite{Amsler:2008zzb}.

\section{Signature of neutral Higgs boson}
\label{HiggsSignature}

One of the clear signal of Higgs boson existence can be seen in
its decay $H \to W^+ W^- \to \mu^+ \nu_\mu + e^- \bar \nu_e$.
Really in final state we will have two different leptons of opposite sign
of electric charge.
In the experiment the energy fractions of leptons and the angle between
the directions of their emission in the system of rest of Higgs boson can be measured.
We show below that this quantities obey non-trivial kinematical
inequality. To obtain this inequality we
consider the phase volume of reaction:
\ba
H \to W^+(q_+) + W^-(q_-) \to (\mu^+(p_+) + \nu_\mu(p_{\nu1})) +
(e^-(p_-) + \bar \nu_e(p_{\nu2})),
\ea
which has the form
$$
\frac{d^3 p_{\nu1}}{2 E_{\nu1}} \frac{d^3 p_{\nu2}}{2 E_{\nu2}}
\frac{d^3 p_+}{2 E_+} \frac{d^3 p_-}{2
E_-}\delta^4(p_H-p_{\nu1}-p_{\nu2}-p_{e1}-p_{e2}) = d^4 p_{\nu1}
\delta\br{p_{\nu1}^2} d^4 p_{\nu2} \delta\br{p_{\nu2}^2}
\times
$$
\ba \frac{d^3 p_+}{2 E_+} \frac{d^3 p_-}{2 E_-}
\delta^4(p_H-p_{\nu1}-p_{\nu2}-p_{e1}-p_{e2}).
\ea
Let's make some transformations in this formula:
\ba
&&
d^4 p_{\nu1} d^4 p_{\nu2}
\to
d^4 q_+ d^4 q_-,
\qquad \br{\mbox{using momenta conservation laws}},\nn\\
&&\frac{d^3 p_+}{2 E_+} \frac{d^3 p_-}{2 E_-}
\to
\frac{1}{4}E_+ dE_+ E_- dE_- d\Omega_+ d\Omega_-,
\ea
where $d\Omega_+$, $d\Omega_-$ are the angular
dependence of final leptons momenta $p_+$ and $p_-$.
Let us now select the pivot direction, say $\vec q_-$,
which will be the initial direction to measure all
the angles from. Then:
\ba
d\Omega_+ d\Omega_- = 2\pi dC_- dC_+ d\phi,
\ea
where $C_\pm = \cos \theta_\pm$ and $\theta_\pm$ are the angles between
$\vec p_\pm$ and $\vec q_-$ and $\phi$ is the angle between the plane
$(\vec q_-, \vec p_+)$ and the plane $(\vec q_-, \vec p_-)$.
So, now we introduce angle $\theta$ (the angle between $p_+$ and $p_-$) and
rewrite the part of phase volume as:
\ba
dC_- dC_+ d\phi =
dC_- dC_+ d\phi dC \delta\br{C-C_+C_-+S_+ S_- \cos \phi} =
\frac{2 dC_- dC_+ dC}{\brm{S_+ S_- \sin\phi}}, \label{DeltaFunc}
\ea
where we introduced new integration over $dC$ ($C=\cos\theta$) with
$\theta$ is the angle between the momenta of the charged leptons (rest frame of
Higgs boson is implied).
The additional $\delta$-function and then we cut off the integration over $\phi$ using
this $\delta$-function.
The denominator $\br{S_+ S_- \sin\phi}$ can be transformed as follows:
\ba
\brm{S_+ S_- \sin\phi}^2 = S_+^2 S_-^2 - S_+^2 S_-^2 C^2 = \nn\\
1 -C_+^2 - C_-^2 - C^2 + 2 C_+ C_- C \equiv D_{H}, \label{Definition}
\ea
where we used the $\delta$-function from (\ref{DeltaFunc}).
After that we use mass-shell $\delta$-functions of neutrinos
$\delta\br{p_{\nu1,2}^2}$ to integrate over $dC_-$ and $dC_+$ in (\ref{DeltaFunc}):
$$
\delta\br{p_{\nu1,2}^2} = \delta\br{\br{q_\mp - p_\mp}^2} =
\delta\br{M_W^2 + m_\mp^2 - 2 E E_\mp \mp 2 \beta E_\mp C_\mp} =
$$
\ba
\frac{1}{2\beta E_\mp} \delta\br{C_\mp - a_\mp}, \ea
where $\beta = \brm{\vec q\pm}/\brm{E_\pm} = \sqrt{1-\frac{4M_W^2}{M_H^2}}$
$m_\pm$ is the mass of corresponding final lepton and
\ba
&&
a=\frac{-M_W^2+2E E_-}{2\beta E E_-} = -\frac{1-\beta^2 - x}{x\beta},
\qquad
x = \frac{4E_-}{M_H}, \label{x}\\
&&
b=\frac{M_W^2-2E E_+}{2\beta E E_+} = \frac{1-\beta^2 - y}{y\beta},
\qquad\qquad
y = \frac{4E_+}{M_H}. \label{y}
\ea
After that the phase volume (\ref{DeltaFunc}) reads as
\ba
\frac{2 dC_- dC_+ dC}{\sqrt{D}},
\ea
which leads to constrain:
$$
D=1 - \br{-\frac{1-\beta^2 - x}{x\beta}}^2 - \br{\frac{1-\beta^2 -
y}{y\beta}}^2 - C^2 +
$$
\ba
2 \br{-\frac{1-\beta^2 - x}{x\beta}} \br{\frac{1-\beta^2 -
y}{y\beta}} C > 0. \ea
So now the size and the shape of this kinematically allowed region depend on the Higgs-mass
(see (\ref{x}), (\ref{y})).
And putting the actual events to 3D-Dalitz plot over three variables ($x$, $y$ and $C$) and fitting
the 3D-body which this events points fill we can find out the mass of Higgs, i.e. the smallest mass of
Higgs which embraces tightly all the measured points by the surface $D=0$.

\section{Discussion}
We have presented a calculated cross section for Higgs boson
production at the LHC, in the decay $H \to WW \to l\nu l\nu$. The
calculation takes into account all the experimental cuts designed
to isolate the Higgs boson signal \cite {CMS, DDDGP}. In the case
of the decay mode $H \to WW \to l\nu l\nu$, we confirm previous
findings that the effect of radiative corrections is strongly
reduced by the selection cuts. \\
We present the spectral distribution on Higgs energy fraction at
some its emission angles.
Our calculations for Higgs boson
production was performed in suggestion of independence of all
vertex functions on the momenta of the particles entering them. We
argue now that this approach is rather realistic. Really the main
mechanisms consists in creation of the jets. The elastic
formfactors contributions are suppressed by Sudakov type
mechanism, whereas the inelastic formfactors or the creation of a
jets are effectively constant of order of ones used above.

We do not consider the two-gamma mechanism of Higgs boson
production as well as it is irrelevant for the large angles
kinematics considered here. The "total" cross section obtained by
the phase volume integration with the reasonable experimental cuts
imposed (the invariant mass of jets of order of proton mass, the
angles between the beam axis and emitted particles as well as
between the final particles supposed to exceed some value
$\theta_i>30^0$ is the quantity of order $1 pb$. From the detailed
analysis of the background situation the Higgs boson
\cite{handbook} decay channel through intermediate two Z- boson
state with the subsequent decay each of them to the pair of
leptons can be identified in data analysis on the level of five
standard deviations (see \cite{handbook}). Further tagging of a pair
of different leptons and the angle between their direction of
motion is described in the previous section. As for the channel
$H\to W^+W^-\to l_1\bar{\nu}_1 \bar{l}_2\nu_2$ the formidable
background appears from the processes with creation of $b\bar{b}$
(see \cite{handbook}).

The accuracy of cross section of light scalar meson production is
estimated on the level of $15-20\%$. It is caused by the frames of
validity of chiral amplitudes method (omission of terms of order
$O(M_p^2/s)$), collinear kinematics approximation and some
uncertainty in the choice of the nucleon formfactors.

\begin{acknowledgments}
The authors wish to thank Prof. M.K. Volkov for fruitful discussions.
Two of us (E.A.K. and Yu.M.B.) acknowledge the
support of INTAS grant no. 05-1000008-8528.
\end{acknowledgments}


\end{document}